# Adaptive Human-Computer Interaction Strategies Through Reinforcement Learning in Complex Scenarios

Rui Liu
*University of Melbourne*
*Melbourne, Australia*

Yifan Zhuang
*University of Southern California*
*Los Angeles, USA*

Runsheng Zhang\*
*University of Southern California*
*Los Angeles, USA*

*Abstract*-This study addresses the challenges of dynamics and complexity in intelligent human-computer interaction and proposes a reinforcement learning-based optimization framework to improve long-term returns and overall experience. Human-computer interaction is modeled as a Markov decision process, with state space, action space, reward function, and discount factor defined to capture the dynamics of user input, system feedback, and interaction environment. The method combines policy function, value function, and advantage function, updates parameters through policy gradient, and continuously adjusts during interaction to balance immediate feedback and long-term benefits. To validate the framework, multimodal dialog and scene-aware datasets are used as the experimental platform, with multiple sensitivity experiments conducted on key factors such as discount factor, exploration rate decay, environmental noise, and data imbalance. Evaluation is carried out using cumulative reward, average episode reward, convergence speed, and task success rate. Results show that the proposed method outperforms existing approaches across several metrics, achieving higher task completion while maintaining strategy stability. Comparative experiments further confirm its advantages in interaction efficiency and long-term return, demonstrating the significant value of reinforcement learning in optimizing human-computer interaction.

## I. Introduction

In the context of rapid digitalization and intelligent development, human-computer interaction has gradually become a key foundation for advancing the information society. From traditional graphical interfaces to voice assistants, virtual reality, and augmented reality systems, interaction modes continue to evolve [1]. The core goal has always been to improve user experience and interaction efficiency. However, as application scenarios grow more complex and user needs become highly personalized, static rules or predefined models are no longer sufficient. Achieving continuous optimization of interaction in complex, uncertain, and open environments has become a major research challenge. Reinforcement learning, with its closed-loop mechanism of trial, feedback, and optimization, offers new possibilities for building adaptive, personalized, and intelligent human-computer interaction systems [2].

Traditional interaction methods often focus on task efficiency and interface design. Such approaches rely on prior experience and manual design, but they lack deep adaptability to user behavior differences and dynamic environmental changes. With the development of artificial intelligence, interaction systems are moving toward greater intelligence and autonomy. The ability to perceive user needs in real time and adjust strategies has become essential. Reinforcement learning, which emphasizes continuous interaction with the environment and learning through reward signals, is well-suited to the optimization of interaction. It enables the exploration of user intentions under uncertainty and dynamic conditions. As a result, it improves efficiency, personalization, and robustness. This makes reinforcement learning one of the most important directions in intelligent human-computer interaction research [3,4].

At the same time, advances in multimodal sensing and computing have extended interaction far beyond single-channel modes. Users often communicate with systems through language, images, gestures, and emotional signals. This complexity increases the demand for experience optimization. The question of how to leverage multimodal information together with reinforcement learning-based decision making to realize adaptive cross-modal strategies is now an important trend. In domains such as large language model interaction [5-7], smart healthcare [8-9], educational support, and industrial information architecture [10-12], user experience directly affects task outcomes and system value. Therefore, reinforcement learning-based optimization of interaction is not only a frontier research problem but also highly relevant for practical applications.

Moreover, optimizing user experience goes beyond convenience and fluency. It also involves trust, satisfaction, and the development of long-term relationships. In intelligent interaction, users expect systems to infer implicit intentions and provide reasonable responses, not merely execute commands. Reinforcement learning can gradually learn user preferences through rewards and embed these preferences into decision strategies. This creates long-term models of individual behavior. Such capabilities give interaction systems a user-centered trajectory, transforming human-computer relationships from tool dependence into intelligent collaboration [13].

In summary, reinforcement learning plays a crucial role in optimizing human-computer interaction. It provides effective solutions for uncertainty and complexity in dynamic environments, overcoming the limits of static modeling. It supports multimodal fusion and personalized modeling, enabling efficient, flexible, and adaptive interaction. More importantly, this line of research promotes the development of intelligent interaction technologies and lays the foundation for applications in education, healthcare, industry, and entertainment [14-16]. Exploring the value and potential of reinforcement learning in human-computer interaction is therefore essential for deep integration between users and intelligent systems and for advancing overall collaboration between humans and machines.

## II. Proposed Approach

In the method design, the overall framework is based on reinforcement learning-based human-computer interaction optimization modeling, abstracting the interaction process between users and the system into a Markov decision process. Its overall architecture is shown in Figure 1.

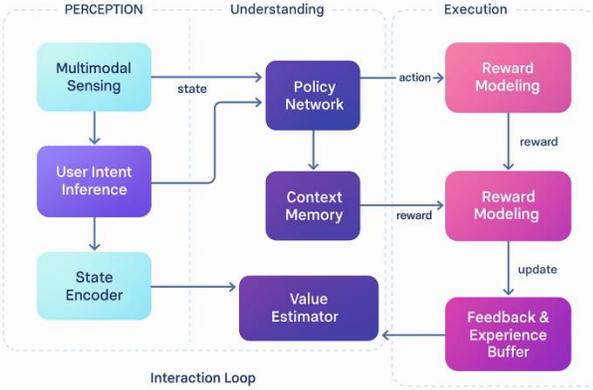

Figure 1. Overall model architecture

Specifically, the system perceives the user's state information at every moment and selects the optimal action based on the policy function, thereby obtaining feedback rewards and updating the policy. This process achieves dynamic optimization of the interactive experience through continuous iteration. Its basic definition can be expressed as a Markov quintuple:

$$M = \{S, A, P, R, \gamma\} \quad (1)$$

Here, S represents the state space, A represents the action space, P represents the state transition probability function, R represents the reward function, and $\gamma \in [0,1]$ represents the discount factor. The state characterizes the user's explicit input and implicit preferences, the action represents the system's interactive feedback, and the reward reflects the positive and negative effects of the user experience.

In the policy modeling process, the system outputs the action selection distribution through the parameterized policy function $\pi_\theta(a|s)$, and combines it with the value function to estimate the long-term return. The value function can be defined as:

$$V^\pi(s) = E_\pi[\sum_{t=0}^{\infty} \gamma^t r_t | s_0 = s] \quad (2)$$

The action-value function further describes the expected return under a given state and action:

$$Q^\pi(s,a) = E_\pi[\sum_{t=0}^{\infty} \gamma^t r_t | s_0 = s, a_0 = a] \quad (3)$$

Based on the above definition, the core goal of interactive optimization is to maximize the cumulative return. To this end, the parameters are updated using the policy gradient method, and the optimization objective function is:

$$J(\theta) = E_{s \sim S, a \sim \pi_\theta}[Q^{\pi_\theta}(s,a)] \quad (4)$$

The gradient update formula is:

$$\nabla_\theta J(\theta) = E_{s, a \sim \pi_\theta}[\nabla_\theta \log \pi_\theta(a|s) Q^{\pi_\theta}(s,a)] \quad (5)$$

On this basis, to enhance the stability and adaptability of the interactive system, the advantage function is introduced to measure the relative advantages and disadvantages of an action compared to the average level. Its definition is as follows [17]:

$$A^\pi(s,a) = Q^\pi(s,a) - V^\pi(s) \quad (6)$$

By introducing the advantage function, the system can more effectively capture the deviation relationship between actions and states, thereby improving the efficiency and convergence speed of policy updates. This modeling approach not only ensures the dynamic and adaptable nature of interactive optimization but also provides a scalable theoretical foundation for personalized modeling of user experience.

## III. Performance Evaluation

### A. Dataset

This study adopts the AVSD (Audio-Visual Scene-Aware Dialog Dataset) as the core dataset. The dataset consists of video clips, audio information, and multi-turn dialogues. It aims to simulate natural interaction scenarios between users and intelligent systems. It covers multiple input modalities, including images, sounds, and textual language. This provides a realistic and complex environment for research on reinforcement learning-driven human-computer interaction. A key feature is that the interaction tasks involve not only simple information retrieval but also scene description, reasoning, and explanation. This allows the dataset to better reflect the diverse demands of real applications.

The dataset is large in scale and contains thousands of videos with complete speech and image information. Each video is paired with multi-turn natural language dialogues. Each dialogue turn reflects the expression of user intent and the dynamic response of the system. This provides abundant training samples for modeling state representation and strategy optimization in human-computer interaction. In particular, the coupling of multimodal signals and temporal sequence features enables support for sequence modeling and long-term dependency learning in complex contexts.

The reason for choosing this dataset lies in its close alignment with real interaction scenarios. It captures the dynamics and uncertainties of user-system exchanges. This characteristic is of great value for the validation and optimization of reinforcement learning methods. It also offers a unified platform for exploring multimodal perception, user intent modeling, and interaction strategy learning. Therefore, the AVSD dataset not only provides a solid data foundation for method design but also shows strong generalizability and application potential.

*B. Experimental Results*

This paper first conducts a comparative experiment, and the experimental results are shown in Table 1.

Table 1. Comparative experimental results

| Method | Cumulative Reward | Average Episode Reward | Convergence Speed | Task Success Rate |
|---|---|---|---|---|
| Mutawa et.al[18] | 215.3 | 10.2 | 180 | 72.4 |
| Ding et. al[19] | 228.7 | 11.5 | 165 | 75.8 |
| Das et. al[20] | 241.9 | 12.3 | 150 | 78.6 |
| Jin et. al[21] | 256.4 | 13.1 | 138 | 81.2 |
| Ours | 289.6 | 14.8 | 110 | 87.3 |

From the overall results, different methods show a gradual improvement in interaction optimization ability, especially in cumulative reward and average episode reward. Traditional methods can improve interaction efficiency to some extent, but they still face limitations in complex and dynamic human-computer interaction scenarios. The proposed method achieves clear advantages in both cumulative reward and average episode reward. This indicates that the method can better capture user intentions and adjust strategies dynamically. As a result, it accumulates higher returns in long-term interactions, highlighting the unique value of reinforcement learning in optimizing user experience.

In terms of convergence speed, the differences between methods further reveal the model's learning efficiency. Early methods often require more iterations to reach a stable state. This leads to higher computational and time costs in real system deployment. The proposed method converges within 110 iterations, which significantly reduces training time compared with other methods. This shows that the model can form stable and effective strategies more quickly. It also demonstrates greater applicability and scalability, making large-scale human-computer interaction systems more feasible.

The improvement in task success rate more directly reflects the effect of interaction optimization. A comparison of methods shows that the proposed method achieves a task success rate of 87.3 percent, which is a clear improvement over traditional approaches. This result indicates that the reinforcement learning framework completes user tasks more efficiently. It reduces redundancy and failures in multi-turn interactions. User goals are achieved more accurately and quickly, which substantially improves the overall interaction quality and verifies the effectiveness of the method in handling complex user demands.

In summary, the proposed method outperforms existing approaches in cumulative reward, learning efficiency, and task completion. More importantly, it presents a scalable approach to optimizing human-computer interaction. Through the dynamic feedback mechanism of reinforcement learning, the system can continuously adapt to changes in both the environment and the user. This enables sustained improvement of the interaction experience. Such capability is of great significance for practical applications. It meets personalized user needs and provides a solid technical foundation for future human-computer collaboration.

This paper further presents an experiment on the sensitivity of the discount factor to the average round reward, and the experimental results are shown in Figure 2.

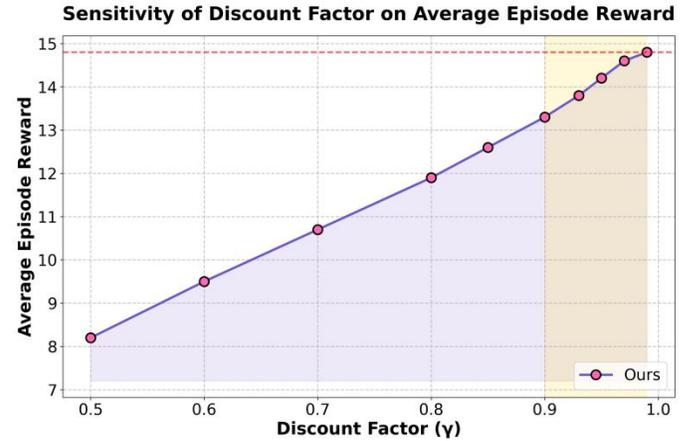

Figure 2. Experiment on the sensitivity of the discount factor to the average round reward

From the overall trend, the average episode reward increases steadily as the discount factor grows. This shows that a higher discount factor guides the model to focus more on long-term returns. In continuous interaction, this leads to better overall performance. The phenomenon matches the core idea of reinforcement learning in human-computer interaction, which is to balance short-term feedback and long-term rewards to improve the quality of experience.

When the discount factor is low, the average episode reward remains at a low level. This indicates that the model tends to emphasize immediate feedback while ignoring long-term optimization. Such a strategy often results in short-term efficiency but fails to sustain user experience over longer sequences. The experimental results show that as the discount factor increases, the system gains a stronger ability to capture user intentions and maintain long-term dialogue coherence. This brings an overall improvement in experience.

Changes in the mid-to-high range reveal that the growth of average episode reward slows and gradually approaches saturation. This means that beyond a certain threshold, the marginal contribution of increasing the discount factor is limited. However, it still helps maintain stable interaction. This feature is important for practical deployment. It suggests that the choice of discount factor should balance convergence speed and long-term benefits to ensure both performance and efficiency.

The final results show that the proposed method achieves the best performance when the discount factor is close to 0.99. This confirms its advantage in modeling long-term rewards. The advantage means that the system can capture implicit user needs more precisely during multi-turn interactions. It can also translate this understanding into stable and efficient strategies. As a result, task success rate and user satisfaction are significantly improved. This finding further validates the potential and theoretical value of reinforcement learning in optimizing human-computer interaction.

This paper also presents a sensitivity experiment on the exploration rate attenuation coefficient to the experimental results, and the experimental results are shown in Figure 3.

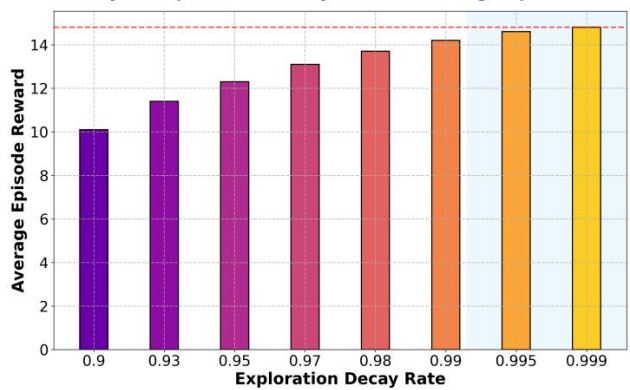

Figure 3. Sensitivity experiment of the exploration rate attenuation coefficient to experimental results

From the overall results, the average episode reward shows a steady upward trend as the exploration rate decay coefficient increases. This indicates that when the system reduces random exploration more quickly and gradually relies on learned strategies, it can effectively improve long-term returns. In other words, a reasonable exploration rate decay mechanism helps the model maintain sufficient diversity in the early stage and later focus resources on strategy optimization. This leads to higher interaction efficiency.

At lower decay coefficients, the average episode reward remains low. This reflects that when the model relies too much on exploration for a long period, strategy exploitation is insufficient. As a result, the cumulative reward during interaction is limited. In this case, the system captures and responds to user intentions with instability, which weakens the overall interaction experience. As the decay coefficient increases, the balance between exploration and exploitation is reshaped. The model focuses more quickly on high-value behavioral patterns, showing stronger performance in understanding user needs and executing tasks.

In the mid-to-high range, the growth of average episode reward slows and gradually reaches saturation. This suggests that when the decay coefficient is too high, the model has already extracted much of the potential of interaction strategies.

Although the improvement at this stage is limited, stability is enhanced. This shows that the system maintains strong adaptability and robustness in complex interaction environments. This is especially important for human-computer interaction, since both overly fast and overly slow exploration decay may cause inconsistent experiences. A reasonable decay schedule ensures the continuous optimization of interaction outcomes.

The final results show that when the decay coefficient is close to 0.999, the average episode reward reaches its highest value. This confirms the advantage of the proposed method in balancing exploration and exploitation. The advantage not only reflects improved interaction efficiency but also demonstrates the ability of reinforcement learning to build long-term stable experiences in human-computer interaction. By dynamically adjusting exploration strategies, the system ensures diverse learning at the beginning and converges more quickly to high-quality user experience paths. This provides a forward-looking optimization direction for human-computer collaboration.

IV. CONCLUSION

This study conducts a systematic investigation into the application of reinforcement learning for optimizing human-computer interaction experience. It proposes an optimization framework that balances long-term returns with immediate feedback. By integrating state representation, policy updating, and reward modeling, the method shows strong capability in capturing user intentions and responding adaptively in complex interaction environments. The results demonstrate that the framework not only improves interaction efficiency but also accumulates higher returns in long-term multi-turn tasks, highlighting the unique value of reinforcement learning in human-computer interaction.

At the methodological level, this study emphasizes the importance of linking policy modeling with user experience evaluation. Through dynamic policy updates and the introduction of an advantage function, the system achieves greater robustness and adaptability in uncertain environments. This approach breaks the limitations of traditional interaction models that focus on single tasks or static scenarios. It also provides a unified optimization idea for cross-modal and multi-task interaction. In particular, under complex conditions involving multimodal information, the proposed method maintains stable performance and offers new directions for the design of intelligent interaction systems.

From an application perspective, the framework presented in this study holds significant value in multiple domains. In education, it can enhance the adaptability of intelligent tutoring systems to student learning states. In healthcare, it can improve the interaction experience in intelligent consultations or rehabilitation support. In industry, it can provide more efficient and safer assistance in complex equipment operations. In entertainment and service applications, it helps create more immersive and personalized interaction processes. These application values indicate that the study goes beyond theory and shows broad practical influence.

In conclusion, this work lays a solid foundation for applying reinforcement learning to the optimization of human-computer interaction. It demonstrates the transferability and practicality of the approach in different application environments. By modeling interaction mechanisms in depth and designing optimization methods systematically, the study verifies that intelligent systems can evolve from tool-based assistance to partner-style collaboration. This transformation is of great importance for deepening human-computer relationships and improving the overall level of interaction systems. It also provides new references and insights for future research in related fields.